\pdfoutput=1 % only if pdf/png/jpg images are used
\documentclass{JINST}
\usepackage[english]{babel}
\usepackage[utf8x]{inputenc}
\usepackage{cite}
\usepackage{graphicx}
\usepackage{multirow}

\title{Performance of the ALICE muon trigger RPCs during LHC Run I}

\author{M. Fontana, for the ALICE Collaboration\\
Università degli Studi di Torino,\\ Italy\\
\\
E-mail: \hspace{0.3cm} \email {mattia.fontana@cern.ch}}

\abstract{ALICE (A Large Ion Collider Experiment) studies the transition of nuclear matter to a deconfined phase known as Quark Gluon Plasma, in ultra-relativistic heavy-ion collisions at the LHC. ALICE is equipped with a muon spectrometer for the detection of quarkonia and heavy flavour particles. The trigger system of the spectrometer consists of 72 RPCs arranged in four detection planes, with a total area of 140 m$\rm ^{2}$. In the first three years of LHC operation, the muon trigger system was fully operational in data-taking in pp, Pb-Pb and p-Pb collisions. The RPC performance and stability throughout the whole data-taking period is presented and discussed, for the parameters such as the efficiency, the dark counting rate, the dark current and the cluster size.}


\begin{document}

\section{Introduction}

ALICE (A Large Ion Collider Experiment\cite{ref::ALICEdet}) collaboration at the LHC studies nuclear matter at very high energy densities, realized in ultra-relativistic heavy-ion collisions. In these conditions, nuclear matter undergoes a phase transition to a deconfined state, known as Quark-Gluon Plasma (QGP\cite{ref::QGP}). The ALICE physics program also includes measurements in proton-proton and proton-nucleus collisions, both of which should serve as reference for the results in nucleus-nucleus collisions, as well as its own interest.\\
From 2009 to 2013 ALICE took data in pp, Pb-p (Pb beam directed towards the muon spectrometer), p-Pb (proton beam directed towards the muon spectrometer) and Pb-Pb collisions under different beam energy and luminosity conditions, summarised in table \ref{tableLumi}.\\
Quarkonium states and heavy-flavour particles are important probes of QGP, their yield being sensitive to the deconfined medium properties; the ALICE muon spectrometer\cite{ref::muonSpectr} is used to detect them via their muonic and semi-muonic decays in the pseudo-rapidity range -4.0$\rm <\eta<$-2.5. The spectrometer is composed of a set of absorbers, a muon tracking system (5 planes of Cathode Pad Chambers), a dipole magnet and a muon trigger system. A detailed description of the detectors and infrastructures is given in\cite{ref::ALICEdet}. The trigger system consists of two stations of two RPC planes each, with 18 RPCs per plane, for a total of 72 chambers. The two stations are set perpendicularly to the beam pipe, at 16 m and 17 m, respectively, from the interaction point. The total active area is about 140~m$^{2}$ and the achieved spatial resolution is better than 1~cm\cite{ref::spatialRes}. The trigger logic exploits the RPC  spatial information; the transverse momentum ($p_{\rm T}$) selection is applied by evaluating the deviation of the muon trajectory with respect to the straight tracks, corresponding to  infinite-momentum case,  originating from the interaction point.\\
The system is able to deliver single and di-muon (unlike- and like-sign) triggers. Two $p_{\rm T}$ thresholds can be handled simultaneously for each of these signals, so that six different triggers in total are evaluated and delivered to the ALICE trigger processor with a latency time of about 800 ns.\\

%The first data taking year was characterised by low luminosity pp collisions at $\rm \sqrt{s}=7$~TeV (8 months/year) and Pb-Pb collisions at $\sqrt{s}=2.76$~TeV (1 month/year). The typical luminosity in pp (Pb-Pb) collisions was about $\rm 10^{29}$~($\rm 10^{25}$)~cm$\rm ^{-2}$~s$\rm ^{-1}$. In 2011, the luminosity reached values of 2$\times$10$^{30}$~cm$\rm ^{-2}$~s$\rm ^{-1}$ for pp collisions at $\sqrt{s}$ = 7 TeV and 5$\times$10$^{26}$~cm$\rm ^{-2}$~s$\rm ^{-1}$ for Pb-Pb collisions at $\sqrt{s_{\rm NN}}$ = 2.76 TeV. Also few pp fills at lower energy ($\sqrt{s}=2.76$~TeV) were taken. 2012 was devoted to pp collisions at $\sqrt{s}=8$~TeV, with a maximum luminosity of $\rm 7\times 10^{30}$~cm$\rm ^{-2}$~s$\rm ^{-1}$. The first months of 2013 were devoted to p-Pb and Pb-p running, with a maximum luminosity of about $\rm 10^{29}$~cm$\rm ^{-2}$~s$\rm ^{-1}$, and to pp collisions at $\sqrt{s}=2.76$~TeV with a maximum luminosity of $\rm 4\times 10^{30}$~cm$\rm ^{-2}$~s$\rm ^{-1}$.

\begin{table}[tbp]
\caption{LHC beam energies and luminosity conditions for the different colliding systems in 2010-2013.}
\label{tableLumi}
\smallskip
\centering
\begin{tabular}{|c|ccc|}
\hline
Year & Colliding system & $ \sqrt{s_{\rm NN}}$ (TeV) & Maximum luminosity (cm$\rm ^{-2}$~s$\rm ^{-1}$) \\
\hline
\multirow{2}{*}{2010} & pp & 7 & $\rm 10^{29}$	\\
 					  & Pb--Pb & 2.76 & $\rm 10^{25}$ \\
 					  \hline
\multirow{3}{*}{2011} & pp & 7 & 2$\times$10$^{30}$\\
 & Pb--Pb & 2.76 & 5$\times$10$^{26}$ \\
 & pp & 2.76 &5$\times$10$^{29}$ \\
 \hline
\multirow{2}{*}{2012} & \multirow{2}{*}{pp} & \multirow{2}{*}{8} & \multirow{2}{*}{$\rm 7\times 10^{30}$}\\
 & & & \\
\hline
\multirow{2}{*}{2013} & p--Pb - Pb--p & 5.02 & $\rm 10^{29}$  \\
 & pp & 2.76 &  $\rm 4\times 10^{30}$\\
\hline
\end{tabular}
\end{table}

The ALICE muon trigger detectors are 2 mm single-gap RPCs, with low resistivity bakelite electrodes ($\rm \sim 10^{9}\,-\,10^{10}\,\Omega\,cm$). The gas mixture consists of $\rm 89.7\,\%\,C_{2}H_{2}F_{4}$, $\rm 10\,\%\,C_{4}H_{10}$, $\rm 0.3\,\%\,SF_{6}$, for a highly saturated avalanche operating mode\cite{ref::ALICEmaxiav}. In this mode, the front-end electronics, called ADULT, discriminates the signals above a very low threshold of 7 mV with no pre-amplification stage\cite{ref::ADULT}. In order to prevent mechanical and chemical alterations to the bakelite electrodes, the mixture is kept at a relative humidity of 37~$\%$\cite{ref::drygas}\cite{ref::Gagliardi:2012zza}.\\
The detectors are read out on both sides with orthogonal copper strips, with pitch of 1 cm, 2 cm and 4 cm and length ranging from 17 cm to 72 cm, in the X (bending plane) and Y (non-bending plane) directions. The operating high voltage ranges from 10 to 10.4 kV. Temperature and pressure variations are corrected using the information given by the online Detector Control System (DCS).\\

\vspace{0.5cm}

In Sec.\ref{sec.2}, the RPC performance in terms of dark rate, dark current, efficiency and cluster size are presented.

\section{RPC performance}\label{sec.2}

\begin{figure}[h!!]
\begin{center}
\includegraphics[scale=0.4]{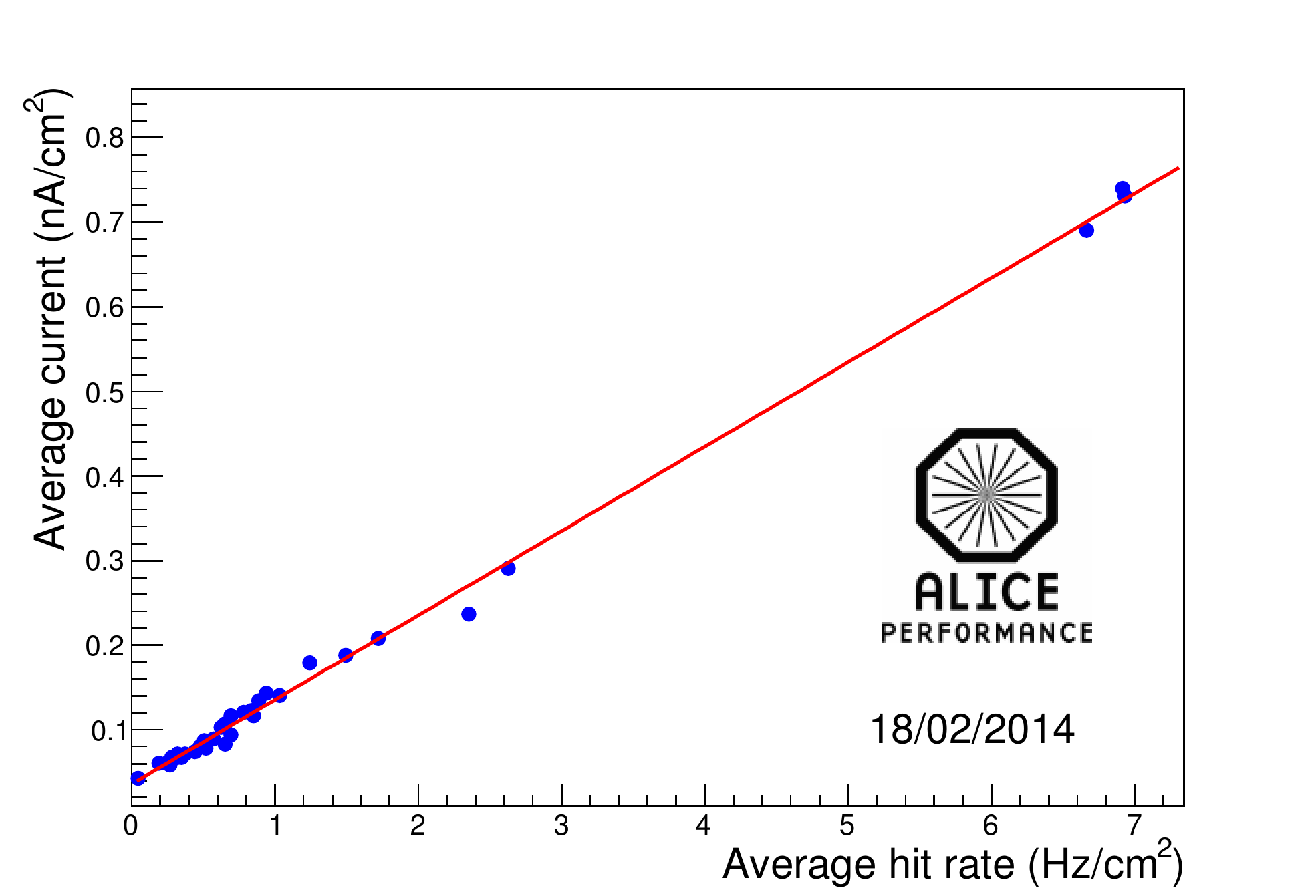}
\caption{Average RPC current versus average counting rate for the pp running mode.}\label{curr-rate}
\end{center}
\end{figure}

During LHC Run I, from 2009 to 2013, the RPC integrated charge was on average 4 $\rm mC/cm^{2}$. The charge for the most exposed detector was 7 $\rm mC/cm^{2}$. The largest contribution from the integrated charge comes from high-rate pp running in 2012. During the R$\&$D phase, RPCs have been successfully tested with the same gas mixture up to an exposure of 50 $\rm mC/cm^{2}$\cite{ref::ALICEmaxiav}\cite{ref::ageingtests}.\\
During Run I, the counting rate on the whole active surface reached values up to 7.5 $\rm Hz/cm^{2}$ for pp, 10 $\rm Hz/cm^{2}$ for Pb-p and 2 $\rm Hz/cm^{2}$ for Pb-Pb.\\
In figure \ref{curr-rate} the average RPC current is shown as a function of the average hit rate: the expected linear correlation is verified, and the fitted slope gives a value of about 100 pC, which corresponds to the average charge associated to a single hit\cite{ref::performance2011}.\\

\subsection{Dark rate}
\begin{figure}[h!]
\centering
%\begin{subfigure}[b]{0.46\textwidth}
\includegraphics[width=0.46\textwidth]{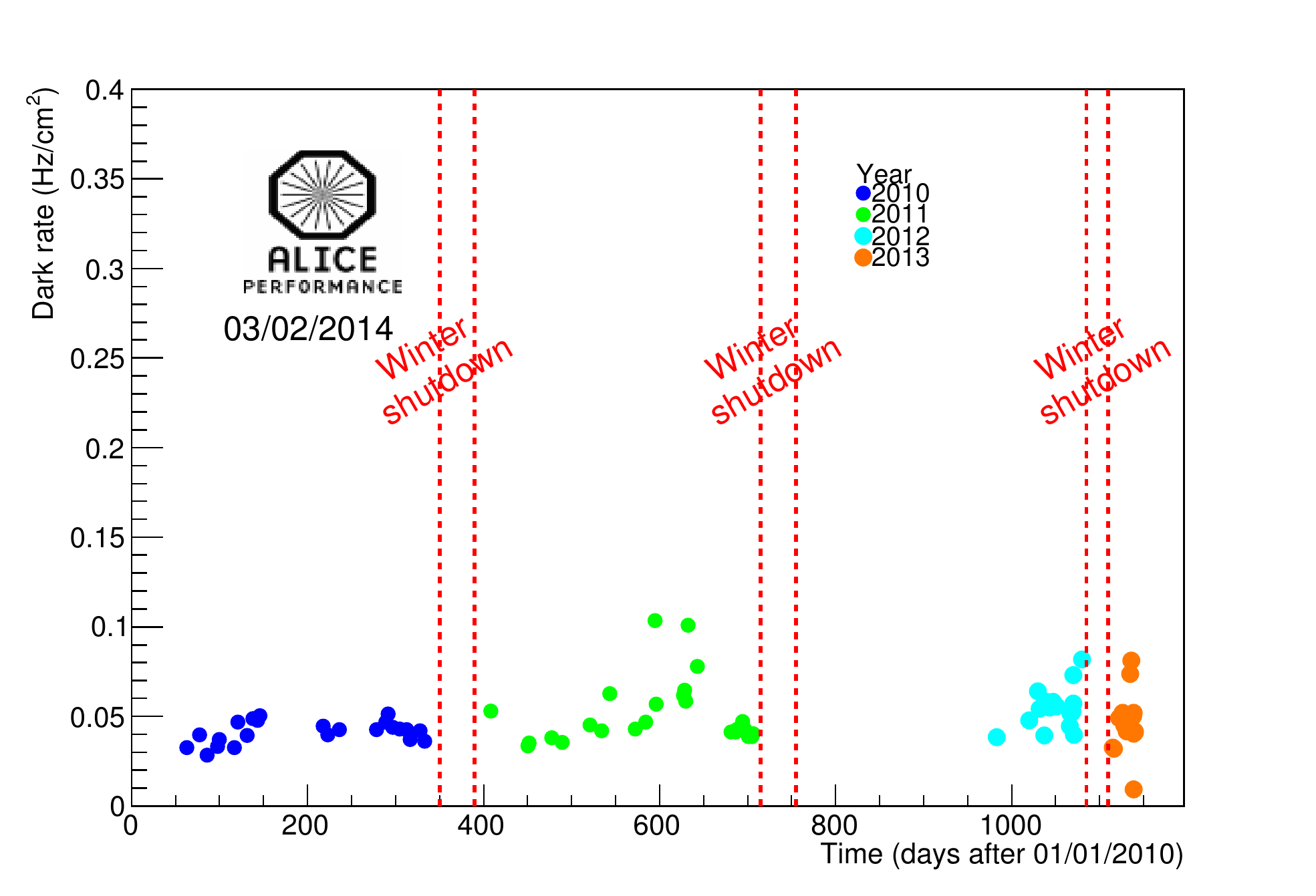}
%\subcaption{Average RPC dark rate as a function of time. \vspace{0.44cm}}
%\label{dark_rate_time}
%\end{subfigure}
\hspace{8mm}
%\begin{subfigure}[b]{0.46\textwidth}
\includegraphics[width=0.46\textwidth]{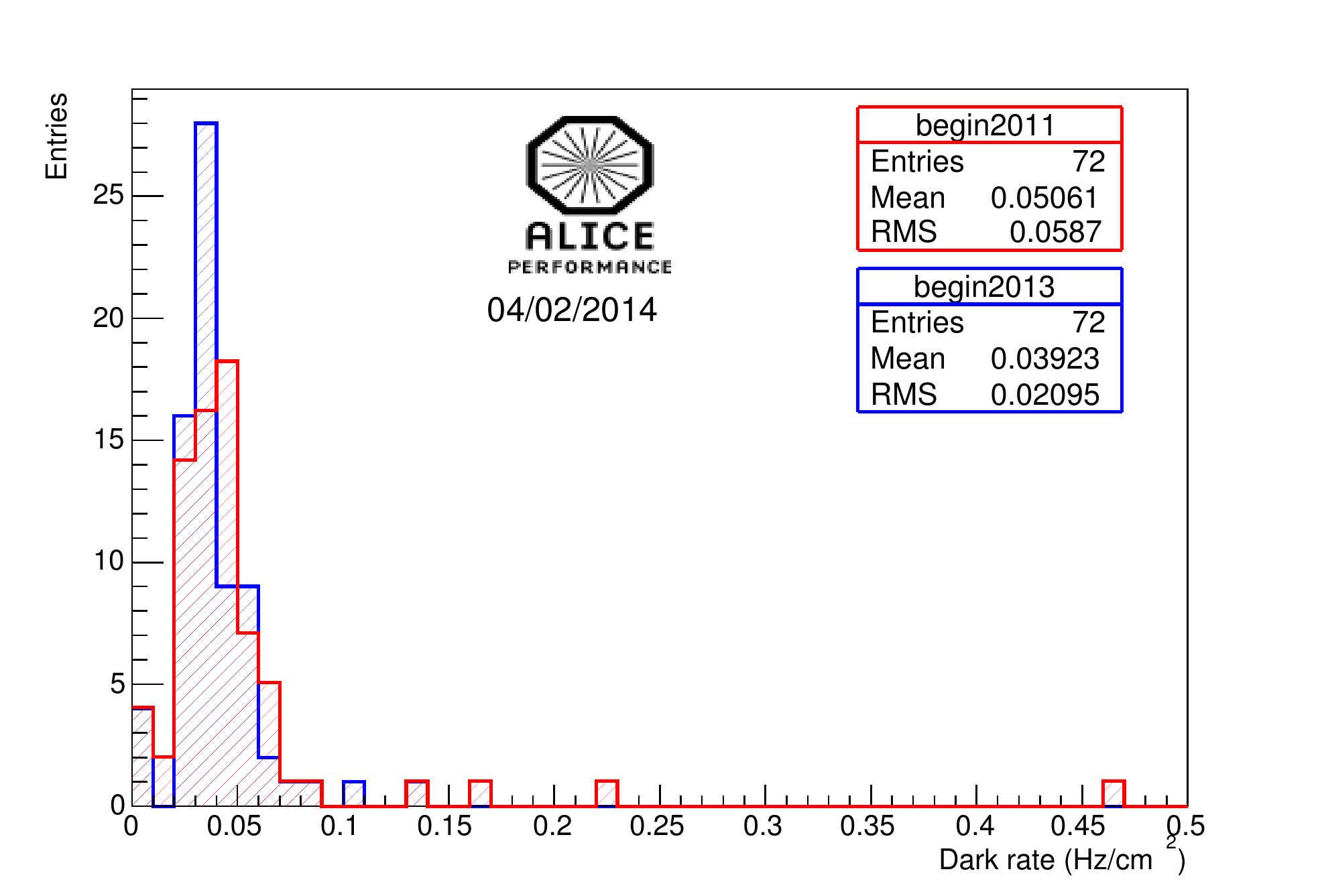}
%\subcaption{Comparison between the RPC dark rate distributions in 2011 (red) and 2013 (blue).}
%\label{dark_rate_distr}
%\end{subfigure}
\caption{Average RPC dark rate as a function of time (left).Comparison between the RPC dark rate distributions in 2011 (red) and 2013 (blue)(right).}\label{dark_rate_fig}
\end{figure}

The dark rate is the detector counting rate with no colliding beams. It is measured in small dedicated runs just after the beam dump. The dark rate has been monitored throughout the whole running period, and the result is shown in figure \ref{dark_rate_fig} (left). A detailed analysis reveals a direct connection between the dark rate value and the features of the previous fill: right after high luminosity fills, the dark rate is higher than after low luminosity ones. Such an effect is most likely due to activation of the surrounding materials. Therefore, only the values measured after low luminosity fills are included in figure \ref{dark_rate_fig} (left).

\vspace{0.8mm}

%\begin{figure}[h!!]
%\begin{center}
%\includegraphics[scale=0.6]{plot/DarkRateVsTime.png}
%\caption{Average dark rate of the four detection planes as a function of time.}
%\label{dark_rate_time}
%\end{center}
%\end{figure}

The dark rate is stable through the whole running period and its average value is always below 0.1 $\rm Hz/cm^{2}$, in line with specifications. The observed stability with time is confirmed by the comparison of the RPC dark rate distribution in 2011 and 2013 (figure \ref{dark_rate_fig} right).

%\begin{figure}[h!!]
%\begin{center}
%\includegraphics[scale=0.4]{plot/RateDistr.png}
%\caption{Dark rate distribution comparison between the beginning of 2011 and the beginning of 2013.}
%\label{dark_rate_distr}
%\end{center}
%\end{figure}
\vspace{0.6cm}

\subsection{Dark current}
\begin{figure}[h!!]
%\begin{subfigure}[b]{0.48\textwidth}
\includegraphics[width=0.48\textwidth]{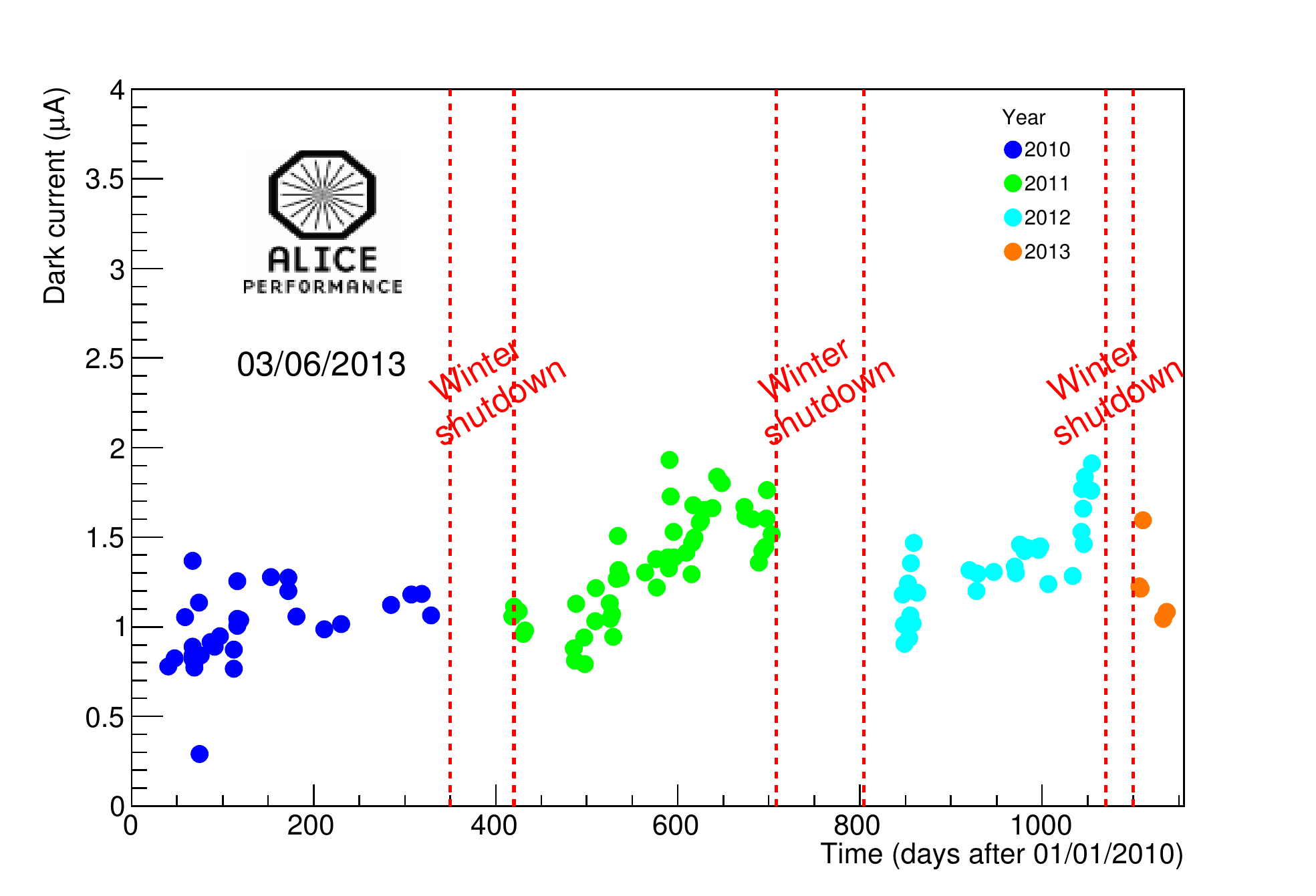}
%\subcaption{Average RPC dark current as a function of time. \vspace{0.05cm}}
%\label{dark_curr_time}
%\end{subfigure}
\hspace{8mm}
%\begin{subfigure}[b]{0.48\textwidth}
\includegraphics[width=0.48\textwidth]{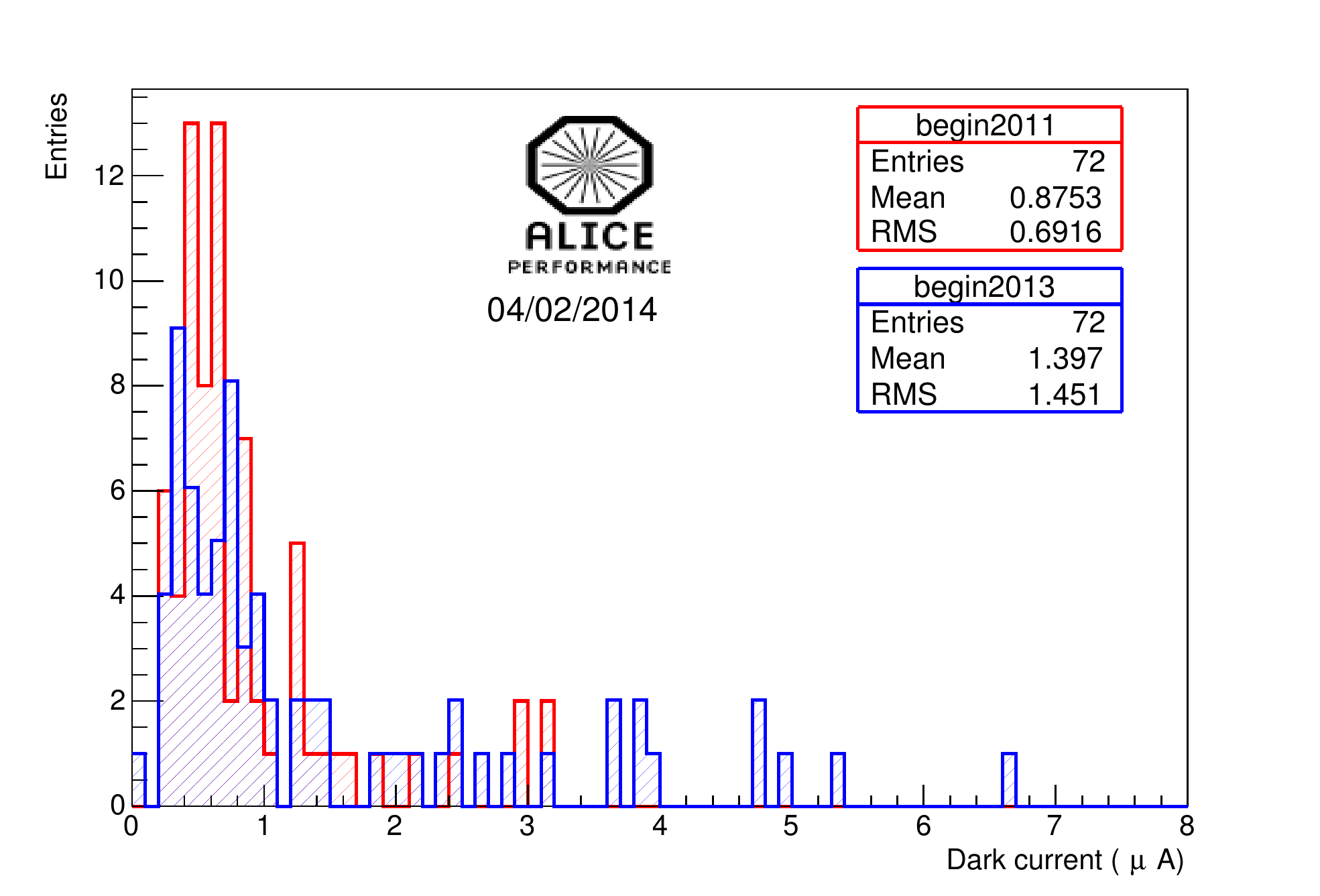}
%\subcaption{Comparison between the RPC dark current distributions in 2011 (red) and 2013 (blue).}
%\label{dark_curr_distr}
%\end{subfigure}
\caption{Average RPC dark current as a function of time (left). Comparison between the RPC dark current distributions in 2011 (red) and 2013 (blue) (right). }\label{dark_curr_fig}
\end{figure}

The dark current has been constantly monitored during the beam-off periods: the evolution of its average value as a function of time is shown in figure \ref{dark_curr_fig} (left).

\vspace{0.8mm}

A slight overall increase of the dark current value is observed during each data-taking year, with a strong decrease after the winter shutdown, during which the detectors were left off with gas flowing. Nevertheless, an overall slight increase of the dark current from year to year is observed. The comparison between the RPC dark current distribution in 2011 and 2013 (figure \ref{dark_curr_fig} (right)) shows that this is due to a number of isolated detectors exhibiting a relatively large increase of the dark current. \\

\begin{figure}[h!]
%\begin{subfigure}[b]{0.52\textwidth}
\includegraphics[width=0.52\textwidth]{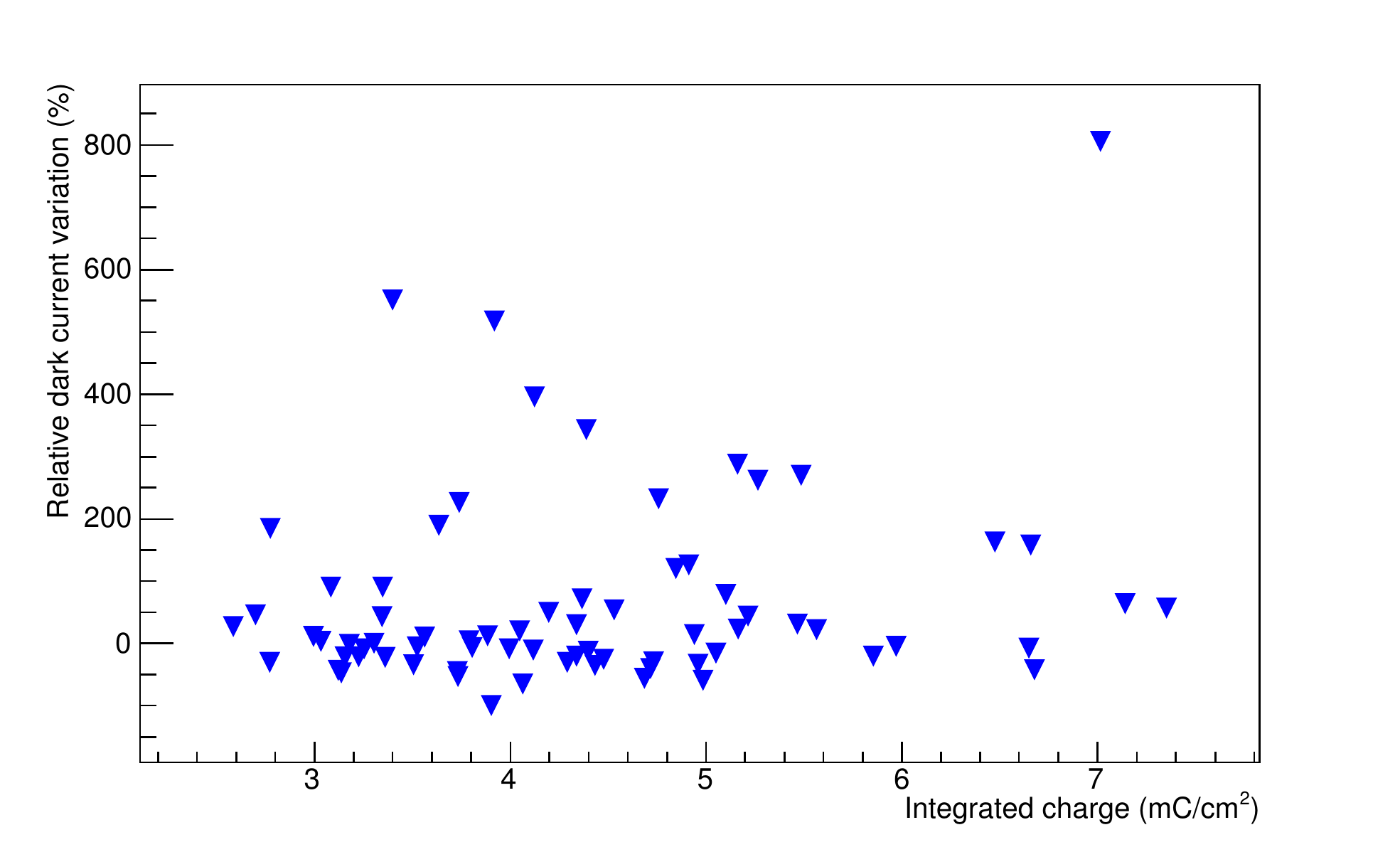}
%\subcaption{Dark current variation between 2010 and 2013 as a function of the integrated charge.}\label{dark_curr_deep_1}
%\end{subfigure}
\hspace{8mm}
%\begin{subfigure}[b]{0.52\textwidth}
\includegraphics[width=0.52\textwidth]{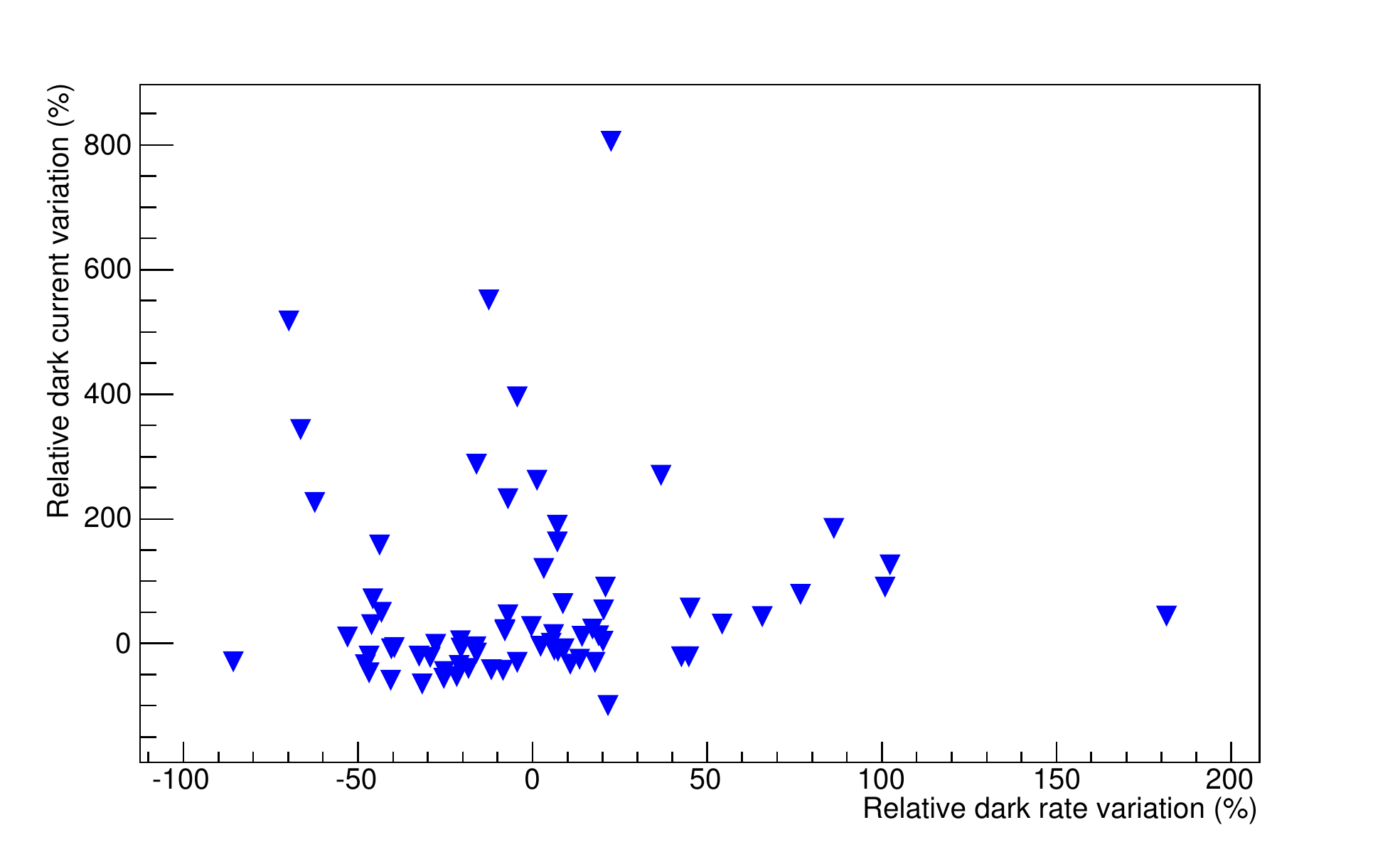}
%\subcaption{Dark current variation between 2010 and 2013 as a function of the dark rate variation.}\label{dark_curr_deep}
%\end{subfigure}
\caption{Dark current variation between 2010 and 2013 as a function of the integrated charge (left). Dark current variation between 2010 and 2013 as a function of the dark rate variation (right). }\label{dark_curr_deep}
\end{figure}

Figure \ref{dark_curr_deep} (left) shows the correlation between the RPC relative dark current increase from 2010 to 2013 and the RPC integrated charge. No evident correlation is observed, indicating that the effect is not related to the irradiation due to beam-beam and beam-gas collisions. Figure \ref{dark_curr_deep} (right) shows the correlation between the dark current increase and the dark rate increase: no evident correlation is observed, indicating that the extra-current is not related to a degradation of the electrode surface.

\vspace{0.8mm}

\subsection{Efficiency}
\begin{figure}[h!!]
\begin{center}
\includegraphics[scale=0.55]{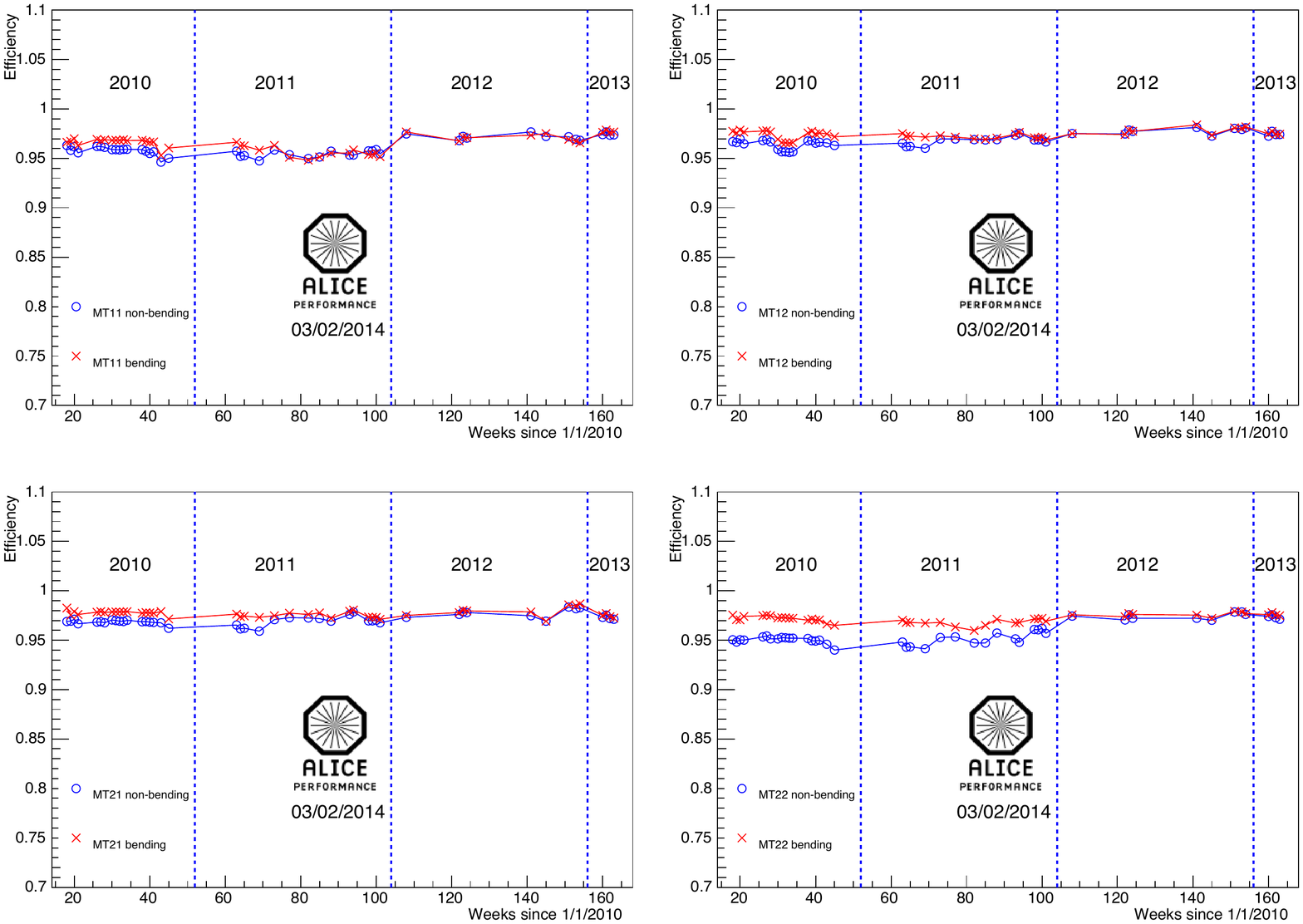}
\caption{Average efficiency of the four detection planes, for both the bending and non-bending planes, as a function of time expressed in weeks since the beginning of 2010.}
\label{Eff_time}
\end{center}
\end{figure}

The RPC efficiency can be measured by exploiting the redundancy of the trigger algorithm, which requires hits in at least three out of four detection planes to give a positive response. Thus, the efficiency of a single detection element can be evaluated by using the remaining three planes as an external trigger and tracking system\cite{ref::efficiency}. In addition to this, being the RPCs equipped with strips on both sides (bending and non-bending planes), the algorithm allows to separately measure the efficiency of the two planes.\\
The efficiency is measured on a weekly basis; in figure \ref{Eff_time} its average value is shown as a function of time for the 4 detection planes.\\

\begin{figure}[htbp]

%\begin{subfigure}[b]{0.47\textwidth}
\includegraphics[width=0.53\textwidth]{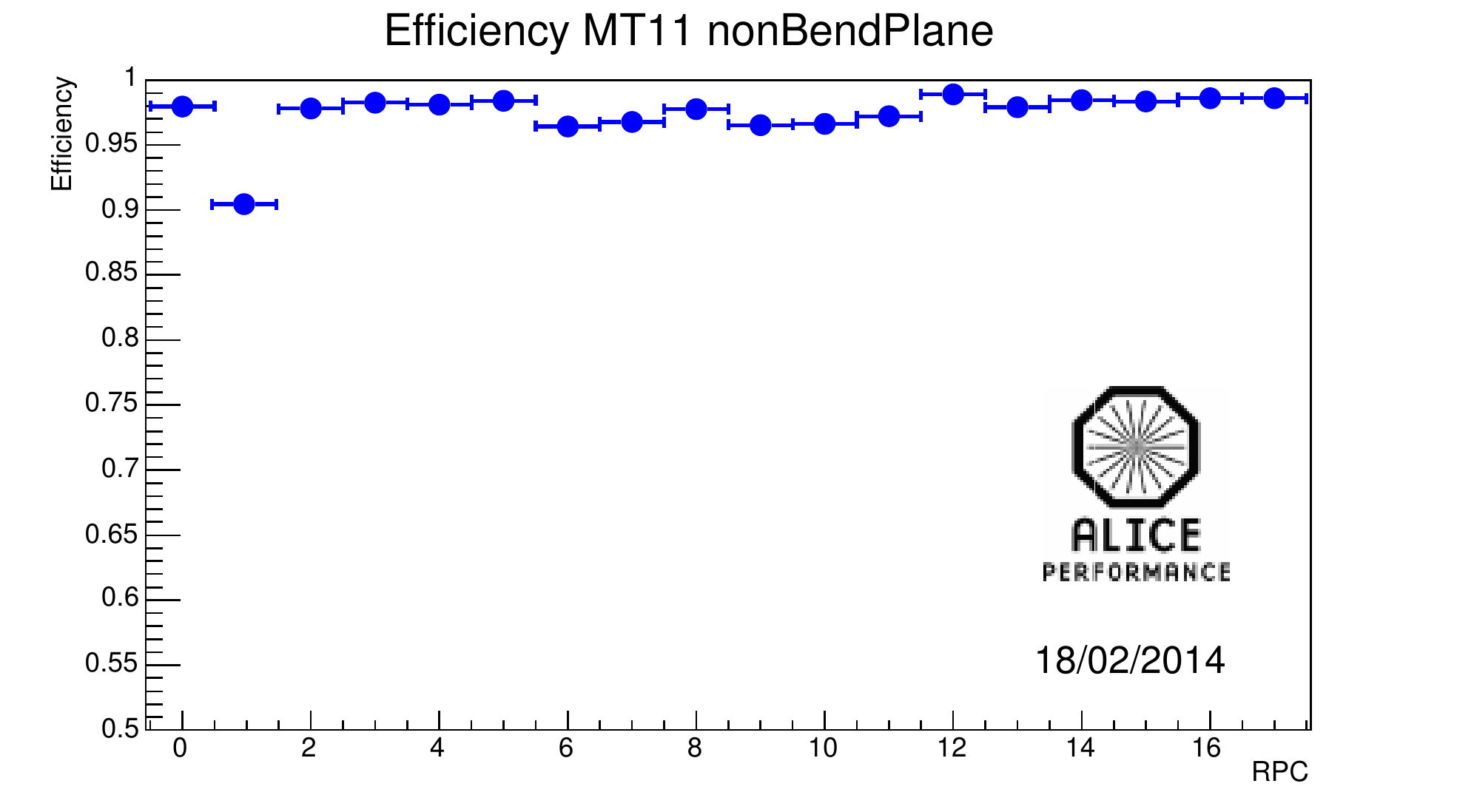}%\label{eff_RPC_MT11}
%\end{subfigure}
\hspace{4mm}
%\begin{subfigure}[b]{0.47\textwidth}
\includegraphics[width=0.53\textwidth]{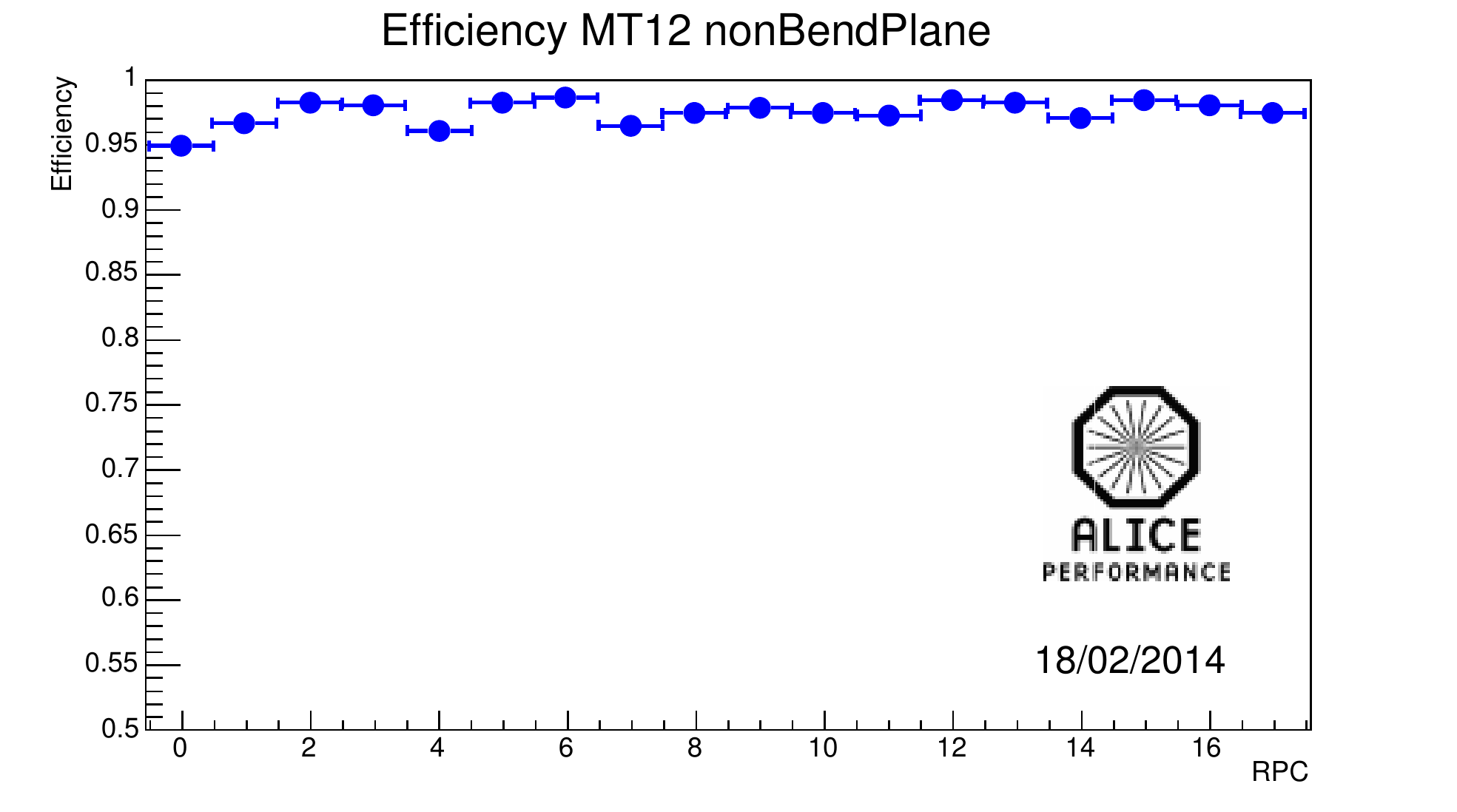}%\label{eff_RPC_MT12}
%\end{subfigure}
%\begin{subfigure}[b]{0.47\textwidth}
\hfill
\includegraphics[width=0.53\textwidth]{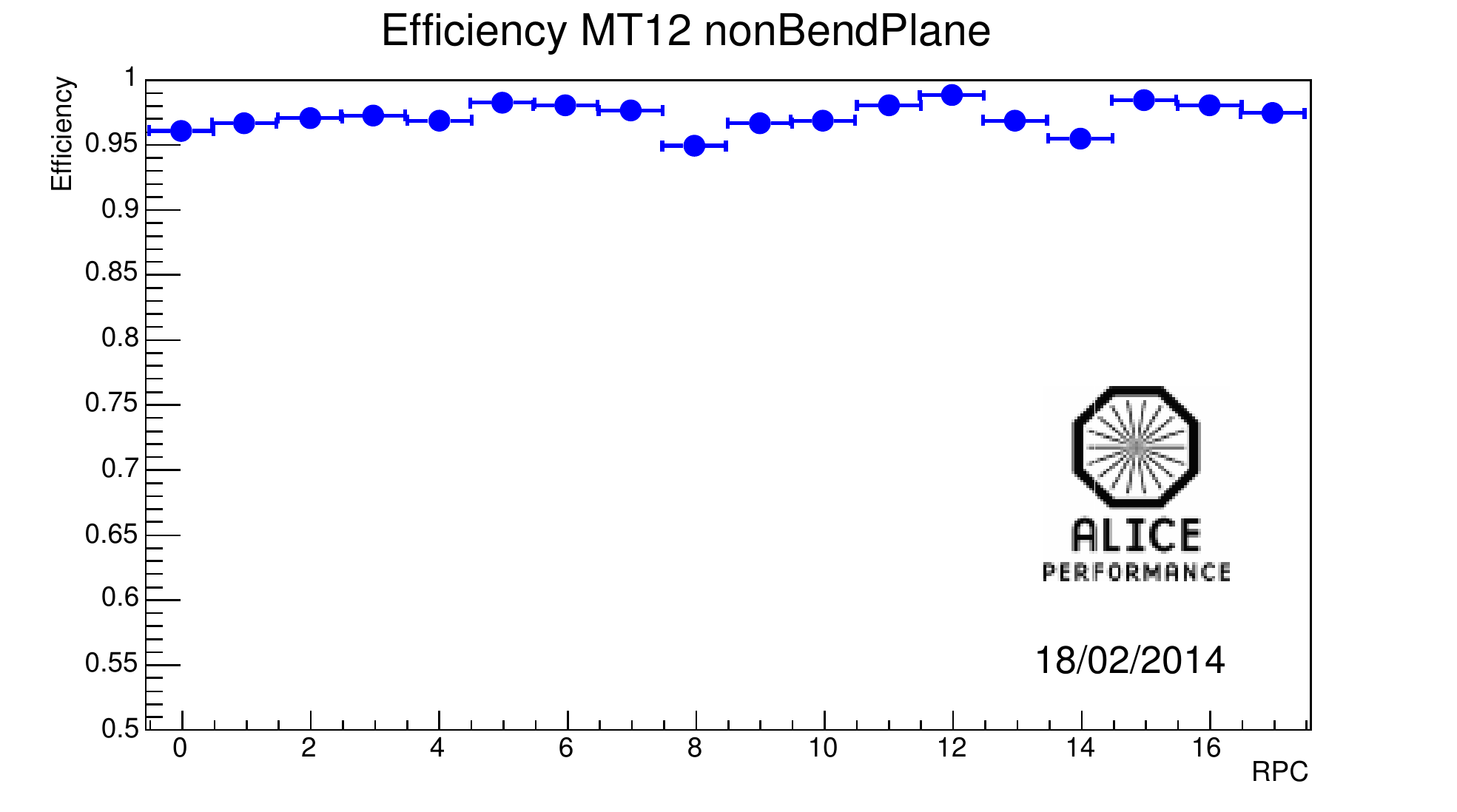}%\label{eff_RPC_MT21}
%\end{subfigure}
\hspace{4mm}
%\begin{subfigure}[b]{0.47\textwidth}
\includegraphics[width=0.53\textwidth]{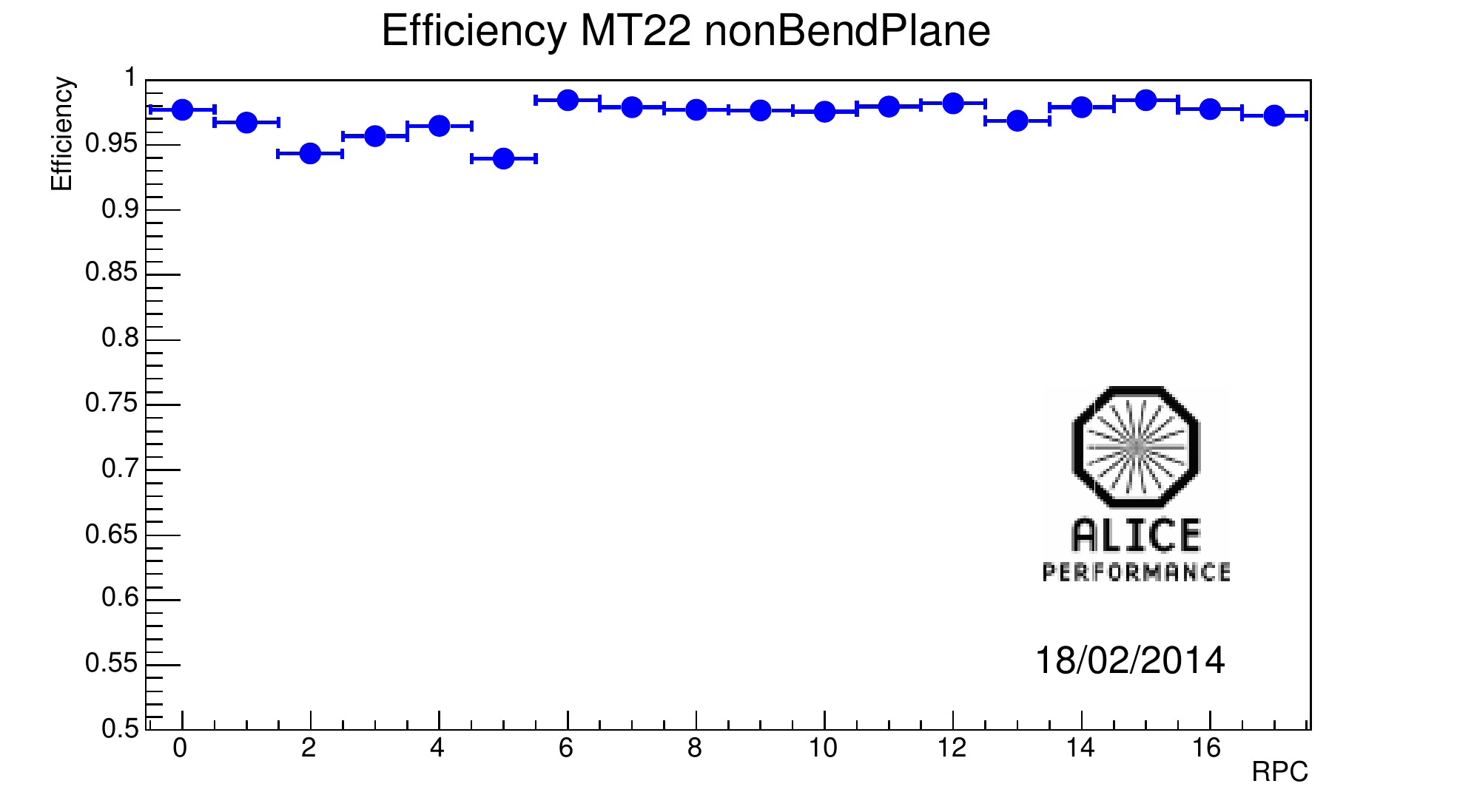}%\label{eff_RPC_MT22}
%\end{subfigure}

\caption{Efficiency of all RPCs at the end of the 2013 data-taking.\label{eff_RPC}}
\end{figure}

The efficiency is stable throughout the whole data-taking period, for the whole RPC system.\\
The individual RPC efficiency has also been studied: all the detectors, with a very few exceptions, have an efficiency larger than 95~$\%$, as can be seen in figure \ref{eff_RPC}.\\

%\begin{figure}[h!!]
%\begin{center}
%\includegraphics[scale=0.35]{plot/eff_RPC.png}
%\caption{Efficiency of all RPCs at the end of the 2013 data-taking.}
%\label{eff_RPC}
%\end{center}
%\end{figure}

\subsection{Cluster size}

\begin{figure}[h!]
%\begin{subfigure}[b]{0.47\textwidth}
\includegraphics[width=0.47\textwidth]{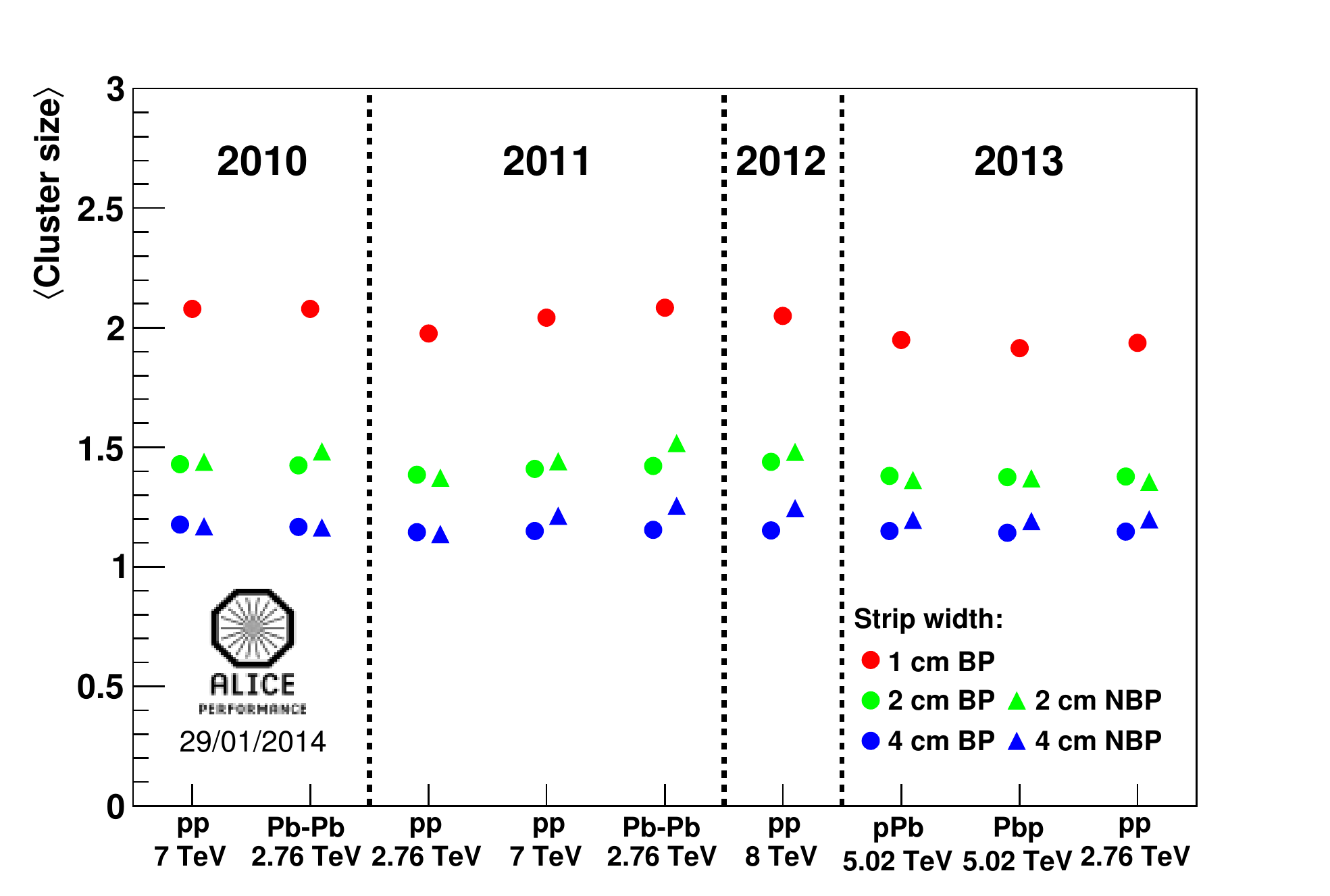}
%\subcaption{Average cluster size during LHC Run I for different colliding systems.}\label{cluster_system}
%\end{subfigure}
\hspace{8mm}
%\begin{subfigure}[b]{0.47\textwidth}
\includegraphics[width=0.47\textwidth]{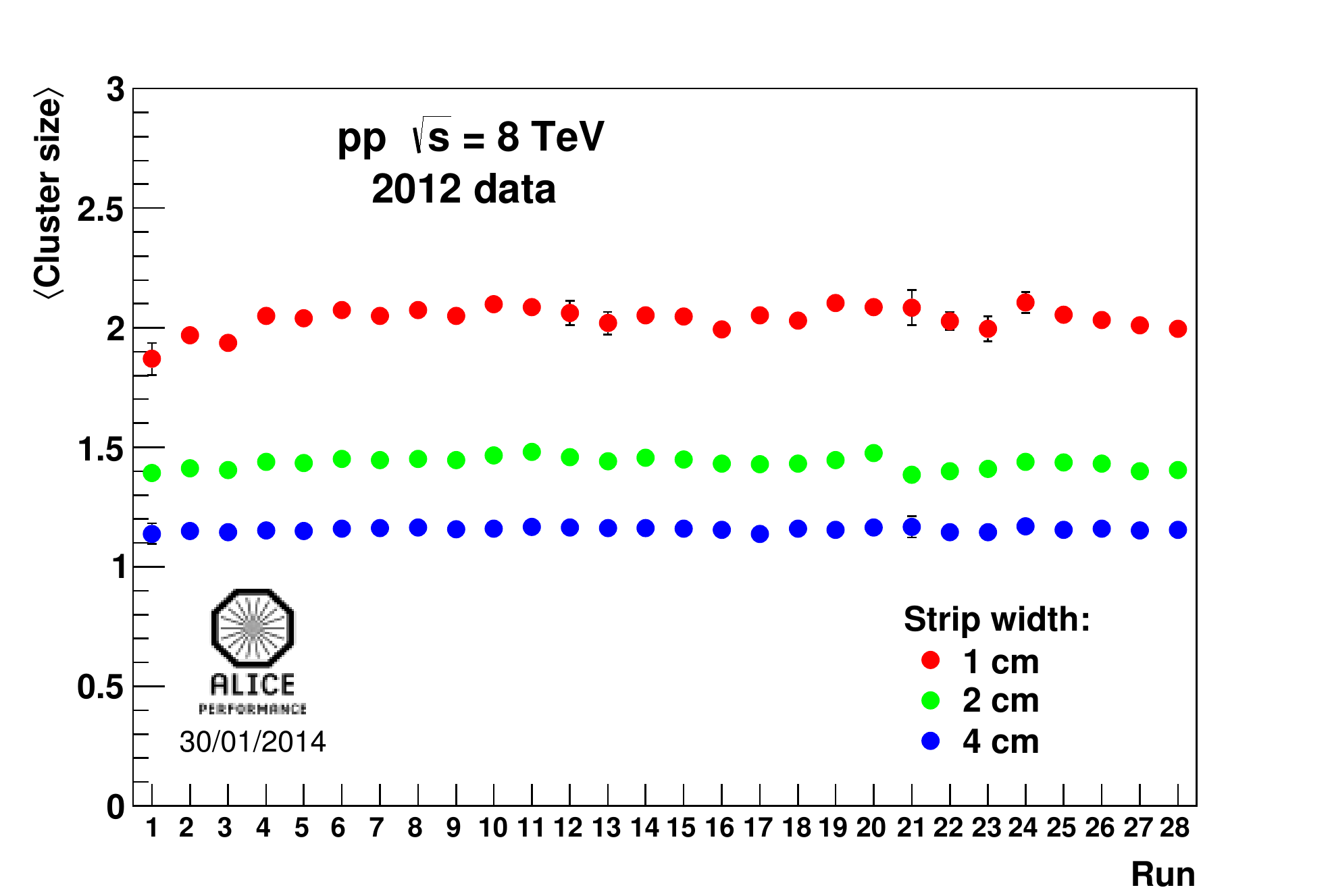}
%\subcaption{Average cluster size as a function of time (expressed as number of runs).}\label{cluster_time}
%\end{subfigure}
\caption{Average cluster size during LHC Run I for different colliding systems (left). Average cluster size as a function of time (expressed as number of runs) (right). }\label{cluster}
\end{figure}

The average cluster size has been measured for all colliding systems during the three years of LHC Run I. In figure \ref{cluster} the average cluster size is shown for all colliding systems (left) and as a function of time during the 2012 pp data-taking (right). The cluster size values are very similar for the different colliding systems, for strips of all pitches. During the R$\&$D phase, the average cluster size was estimated as 1.3 for 2 cm strips with a 10 mV threshold\cite{ref::ALICEmaxiav}. Despite the different threshold (7 mV at run time), the results obtained in the runs are very close to the R$\&$D ones.\\

\vspace{0.8mm}

A satisfactory stability in time is also observed (figure \ref{cluster} right).

\subsection{Trigger efficiency}
\begin{figure}[h!!]
\begin{center}
\includegraphics[scale=0.35]{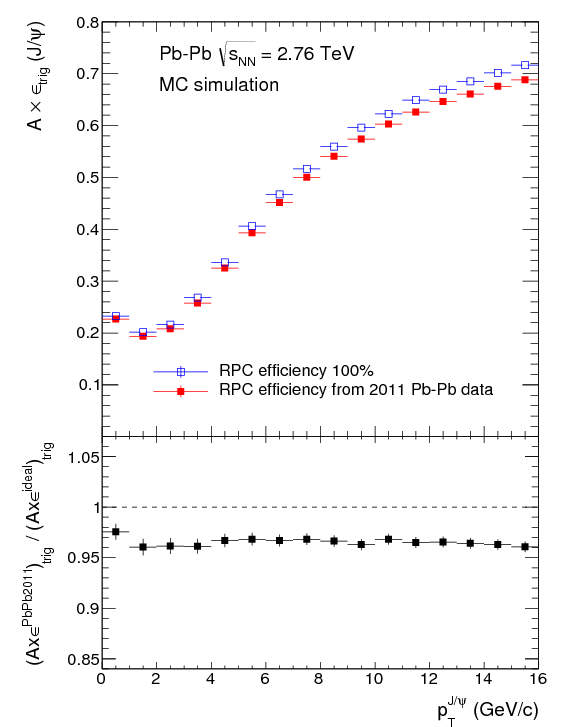}
\caption{RPC trigger efficiency for $\rm{J}/ \psi$ detection as a function of transverse momentum for ideal (100$\%$ efficiency) and real case (efficiency extracted from data). The ratio between the two simulations is also shown. This figure is taken from Ref.\cite{ref::triggEff}.}
\label{trigg_eff}
\end{center}
\end{figure}

Figure \ref{trigg_eff} shows the muon trigger efficiency for $\rm{J}/ \psi$ detection: the ideal case (RPC 100$\%$ efficiency) and the real one (with the efficiency extracted by Pb-Pb data from 2011) are compared as a function of the $\rm{J}/ \psi$ transverse momentum. The ratio between the two is also shown. This is an example of how the detector performance translates into physics performance.

The comparison between the real and the ideal case is satisfactory, the two being always within 5$\%$.
The effect of the trigger chamber inefficiencies is smaller than 5$\%$, with weak (if any) $p_{\rm T}$ dependence.

\section{Conclusions}

The ALICE muon trigger system has been fully operational during LHC Run I. The observed RPC performance is in agreement with the design value. The average RPC efficiency is typically larger than 95$\%$ and stable in time; the dark rate is stable and always below 0.1 $\rm Hz/cm^{2}$. The dark current shows a slight overall increase during data taking periods followed by a decrease after long winter stops. A few RPCs show a systematic increase of the dark current, not correlated with the integrated charge and not justified by an increase of the dark rate. The average cluster size is about 1.4 for 2 cm strips, in line with specifications.\\
\\
The ALICE muon spectrometer is playing a crucial role in the ALICE physics program and it will continue to do so also during the LHC Run II, without hardware modifications. An upgrade for Run III has been  planned and is now being carried out\cite{ref::baptiste}. The detectors will be operated in genuine avalanche mode in order to cope with the harsher running conditions expected\cite{ref::ALICELOI}.

%\acknowledgments

%\begin{thebibliography}{9}
%\bibliographystyle{unsrt}
%\bibliography{biblio}{}

%\end{thebibliography}
\end{document}